\documentclass{epl}

\usepackage{amsmath,amssymb,amsfonts}
\usepackage{graphicx,subfigure}
\usepackage{cite}
\usepackage{graphics}

\newcommand{\avg}[1]{\left\langle{#1}\right\rangle}

\newcommand{\ovl}[1]{\overline{#1}}

\title{Adaptive drivers in a model of urban traffic}
\author{A. De Martino\inst{1} \and M. Marsili\inst{2} \and
R. Mulet\inst{3}} \institute{ \inst{1} INFM--SMC and Dipartimento di
Fisica, Universit\`a di Roma ``La Sapienza'', p.le A. Moro 2, 00185
Roma (Italy)\\ \inst{2} The Abdus Salam ICTP, Strada Costiera 14,
34014 Trieste (Italy) \\ \inst{3} Henri Poincar\'e Chair of Complex
Systems and Superconductivity Laboratory, Physics Faculty and IMRE,
Universidad de la Habana, CP 10400 La Habana (Cuba)}

\begin{document}

\maketitle

\vspace{-0.5cm}

\begin{abstract}
We introduce a simple lattice model of traffic flow in a city where
drivers optimize their route-selection in time in order to avoid
traffic jams, and study its phase structure as a function of the
density of vehicles and of the drivers' behavioral parameters via
numerical simulations and mean-field analytical arguments. We identify
a phase transition between a low- and a high-density regime. In the
latter, inductive drivers may surprisingly behave worse than randomly
selecting drivers.
\end{abstract}

\vspace{-0.5cm}

After the seminal works \cite{1,2,3}, models of vehicular traffic have
enjoyed a continuously increasing interest among physicists (see
e.g. \cite{hg,pr,rmp}), and substantial progress has been achieved in
understanding the origin of many empirically observed features. In
several cases, traffic models have also revealed connections to
important out-of-equilibrium systems in statistical mechanics. The
NaSch cellular automaton \cite{2,ss}, for instance, is a close
relative of the totally asymmetric simple exclusion process (TASEP)
\cite{derrida}, and of the KPZ-class of surface growth models
\cite{kpz,hh}. Urban traffic has also been extensively
studied. Examples are BML models \cite{1,pr}, where vehicles are bound
to travel on a lattice, representing the network of streets, with
time-evolution governed by TASEP-like rules. Typically, increasing the
density of vehicles, a sharp transition occurs from an unjammed regime
with finite average velocity to a jammed state where cars are blocked
in a single large cluster spanning the entire lattice.

In these cellular-automaton type of models, the dynamics of drivers
does not pursue an explicit goal, like minimizing traveling
times. Here we introduce a different class of models. We postulate
that each driver has a finite set of feasible routes for going between
two points in a city and that, on each day, he/she tries to choose the
least crowded one using a simple learning process \cite{helbo}. Our
principal aim is to investigate whether -- and in which traffic
conditions -- inductive drivers behave more efficiently than drivers
who choose their routes at random. A similar question arises in the
Minority Game (MG) \cite{cz1}, and this affinity constitutes the
starting point of our analysis. More precisely, upon increasing the
density of vehicles we find a transition from a phase where drivers
are unevenly distributed over streets to one where streets are equally
loaded. In this latter phase, depending on their learning rates,
inductive drivers can behave worse than randomly selecting drivers
because of crowd effects.

\begin{figure}
\begin{center}
\includegraphics[width=4.5cm]{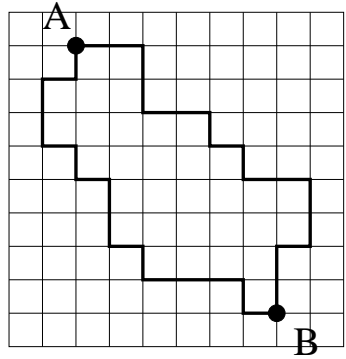}
\caption{\label{strat}Two routes of length $\ell=16$ for going from
$A$ to $B$ in a $11\times 11$ city.}
\end{center}
\end{figure}

We consider, as a model of a city, a square lattice with $L\times L$
nodes and $P=2L^2$ edges (assuming periodic boundary conditions). Edges
represent the different streets.  On each day, each one of $N$ drivers
must travel from a node $A$ (say, his/her workplace) to a node $B$
(say, home) ($A$ and $B$ are different for different drivers)
following one of $S$ possible routes. Routes are sets of consecutive
edges on the lattice (see Fig.~\ref{strat}).
Indexing edges by $\mu$, it is natural to identify a route with a
vector $\boldsymbol{q}_{ig}=\{q_{ig}^\mu\}_{\mu=1}^P$, where
$i=1,\ldots,N$ labels drivers and $g=1,\ldots,S$ labels the feasible
paths of each driver, while $q_{ig}^\mu$ is the number of times driver
$i$ passes via street $\mu$ in route $g$. For each $i$, $S$ randomly
generated routes of equal length $\ell$ (length $=$ number of edges)
joining two randomly chosen nodes $A$ and $B$ are given\footnote{In
our simulations, loops are admitted, but all results remain valid for
more realistic loop-less routes.}. Denoting by $\widetilde{g}_i(n)$
the route selected by driver $i$ on day $n$, we assume that
\begin{equation}\label{logit}
{\rm Prob}\{\widetilde{g}_i(n)=g\}=C~e^{\Gamma U_{ig}(n)}
\end{equation}
with $C$ a normalization constant and $\Gamma\geq 0$ the `learning
rate' of drivers. The functions $U_{ig}$ are the scores of the
different feasible routes of driver $i$, and are updated according to
\begin{equation}\label{id}
U_{ig}(n+1)=U_{ig}(n)-\frac{1}{P} \sum_{\mu=1}^{P} q_{ig}^\mu
Q^{\mu}(n)+\frac{1}{2}(1-\delta_{g,\widetilde{g}_i(n)})\zeta_{ig}(n), 
\end{equation}
where $Q^\mu(n)=\sum_{i=1}^N q_{i,\widetilde{g}_i(n)}^\mu$ is the
number of drivers passing through $\mu$ on day $n$. We assume this to
be known to every driver, e.g. via the GPS technology or the traffic
bulletin, to a degree specified by the last term on the r.h.s. of
(\ref{id}), which accounts for the fact that the driver's knowledge of
the $Q^\mu(n)$ of routes he/she has not visited on day $n$
(i.e. streets such that $q_{i\widetilde{g}_i(n)}^\mu=0$) is subject to
the noise $\zeta_{ig}(n)$, with mean value and correlations given by
\begin{equation}
\langle \zeta_{ig}(n)\rangle=\eta ~~~~~~~~~~~~~
\langle \zeta_{ig}(n) \zeta_{jh}(m)\rangle=
\Delta\delta_{ij}\delta_{gh}\delta_{nm}
\end{equation}
If $\eta=0$, the driver has unbiased information about routes he/she
does not follow, whereas if $\eta>0$ (resp. $\eta<0$) he/she
overestimates (resp. underestimates) the performance of such
routes. We shall first consider the simpler case $\eta=\Delta=0$, and
later discuss the effects introduced by $\eta\neq 0$ and $\Delta>0$.

Altogether, (\ref{logit}) and (\ref{id}) imply that drivers prefer
less crowded routes. If one assumes that the transit time through
street $\mu$ is proportional to $Q^\mu$, then the above dynamics
describes a driver trying to learn which of his feasible routes is
faster. For $\Gamma\to\infty$, drivers always choose the route with
the highest score, whereas for $\Gamma<\infty$ they randomize the
choice. In particular, for $\Gamma=0$ they select one of their
feasible routes at random with equal probability.

\begin{figure}
\begin{center}
\includegraphics[width=7.5cm]{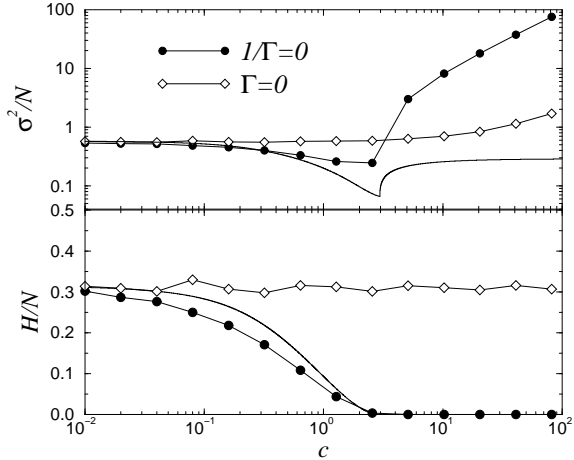}
\caption{\label{traffic}Stationary values of $\sigma^2/N$ and $H/N$
as a function of $c=N/P$ for randomly selecting drivers ($\Gamma=0$,
open markers) and adaptive drivers ($\Gamma=+\infty$, filled
markers). In these simulations: $S=2$, $P=200$, $\ell=50$ and averages
are taken over at least $50$ disorder samples for each point. The
initial conditions of (\ref{id}) were taken to be $U_{ig}(0)=0$ for
all $i$ and $g$. The solid line in the graphs is the analytic estimate
of $H$ and $\sigma^2$ (for $\Gamma= 0^+$) for the model with
uncorrelated disorder.}
\end{center}
\end{figure}

As in the MG~\cite{Savit}, we expect the collective performance in the
stationary state of (\ref{id}) to depend only on the density of
vehicles $c=N/P=N/(2L^2)$. We focus on the asymptotic (long-time)
values of the observables \cite{cmz}
\begin{equation}
H=\ovl{\avg{Q^\mu}^2}-\ovl{\avg{Q^\mu}}^2
\qquad\text{and}\qquad
\sigma^2=\ovl{\avg{(Q^\mu)^2}}-\ovl{\avg{Q^\mu}}^2
\end{equation}
where $\avg{\cdot}$ stands for a time average over the stationary
state of (\ref{id}) and $\ovl{x^\mu}=(1/P)\sum_{\mu=1}^{P} x^\mu$. $H$
describes the distribution of drivers over the street network in the
stationary state. If $H=0$, the distribution is uniform
($\avg{Q^\mu}=\ovl{\avg{Q^\mu}}$ for all $\mu$) and it is not possible
to find convenient routes. If $H>0$, instead, the distribution is not
uniform and fast pathways do exist. The optimal road usage is achieved
when $H=0$ and traffic fluctuations (namely $\sigma^2-H$) are
minimized. Notice that if transit times are assumed to be proportional
to the street loads $Q^\mu$, then $\sigma^2$ measures the total
traveling time of drivers. Notice also that, since all routes have the
same total length $\ell$ (i.e. $\sum_\mu q_{ig}^\mu=\ell$ for all $i$
and $g$), $\ovl{\avg{Q^\mu}}=c\ell$ is just a constant.  In
Fig.~\ref{traffic}, we report the results obtained from computer
experiments with $\eta=\Delta=0$ as a function of $c$ for drivers with
$\Gamma=\infty$ and $\Gamma=0$.
The latter lead to a stationary state where a uniform distribution of
vehicles is never achieved, as for them $H>0$ for all $c$. The former,
instead, behave in a similar way only for small $c$. As $c$ is
increased, the traffic load becomes more and more uniform ($H$
decreases) and fluctuations ($\sigma^2$) decrease, indicating that
inductive drivers manage to behave better than random ones. At a
critical point $c_c\simeq 3$ the distribution becomes uniform
(i.e. $H=0$) and vehicles fill the available streets uniformly. Now
drivers can't find a convenient way and are forced to change route
very frequently. As a consequence, global fluctuations increase
dramatically. Notice that above the critical point traffic
fluctuations are significantly smaller for randomly selecting drivers
than for optimizers.

This behavior is very similar to that of the MG. It is easy to check,
along the lines of \cite{mc}, that in the stationary state drivers
choose routes with frequencies $f_{ig}=\avg{{\rm
Prob}\{\widetilde{g}_i(n)=g\}}$ that minimize the `Hamiltonian'
\begin{equation}\label{acca}
{\cal H}_\eta(\{f_{i,g}\})=\ovl{\avg{Q^\mu}^2}+\frac{\eta}{2}\sum_{i,g}
f_{ig}(1-f_{ig}/2), ~~~~~\avg{Q^\mu}=\sum_{i,g} f_{ig}q_{ig}^\mu
\end{equation}
subject to $\sum_g f_{ig}=1$. In fact, taking the average of
(\ref{id}) in the stationary state, one sees that the stationarity
conditions for $f_{ig}$ are exactly the conditions for the minima of
${\cal H}_\eta$. Note that, up to a constant, ${\cal H}_0=H$. The
stationary frequencies $f_{ig}$ turn out to be independent of $\Gamma$
for $\Gamma>0$ (see \cite{mc} for details) and of $\Delta$. In turn,
because of (\ref{acca}), $H$ has the same property.

A direct application of the standard replica-based minimization tools
to compute $\min H$ is made difficult by the presence of spatial
correlations\footnote{If a route goes through street $\mu$
($q_{ig}^\mu>0$) then nearby streets are more likely to belong to the
same path than streets which are far away.}  in the disorder variables
$q_{ig}^\mu$.  In order to test the effects of spatial correlations,
we considered a `na\"\i ve' version of (\ref{id}), in which the
$q_{ig}^\mu$'s are i.i.d. quenched random variables with the same
probability distribution as in the lattice setup. The minima of $H$
for this simplified model can be easily studied with the replica
method (we do not report this analysis here). It turns out (see
Fig.~\ref{traffic}) that the shape of the functions $H$ and $\sigma^2$
(for small but non-zero $\Gamma$, i.e. $\Gamma=0^+$, see below) are
qualitatively the same as those of numerical simulations at
$\Gamma=\infty$. The critical point $c_c= 2.9638\ldots$ computed
analytically is very close to that found in numerical simulations.

The knowledge of $f_{ig}$, however, does not allow us in principle to
compute $\sigma^2$, which requires a fully dynamical
approach. Following \cite{mc}, we shall now derive the continuous-time
(stochastic) differential version of (\ref{id}). To begin with, we
restrict ourselves to the case $S=2$ ($g=1,2$) and introduce the
variables $y_i(n)=[U_{i1}(n)-U_{i2}(n)]/2$ and
$s_i(n)=3-2\widetilde{g}_i(n)$, such that $s_i(n)=\pm 1$ is an Ising
spin. The dynamics can then be re-cast in the form
\begin{equation}\label{manna}
y_i(n+1)=y_i(n)-\frac{1}{P}\sum_{\mu=1}^P\xi_i^\mu Q^\mu(n)
+\frac{1}{8}\left[(1-s_i(n))\zeta_{i1}(n)-(1+s_i(n))\zeta_{i2}(n)\right]
\end{equation}
where $\xi_i^\mu=(q_{i1}^\mu-q_{i2}^\mu)/2$. Simulations show that
relaxation and correlation times are proportional to
$1/\Gamma$. Following \cite{mc}, we introduce the re-scaled time
$t=\Gamma n$ and study the dynamics in a time interval $\Delta t$
corresponding to $\Delta n=\Delta t/\Gamma$ time steps. When
$\Gamma\ll 1$, we can take $\Delta n$ large while keeping $\Delta t$
small. A large $\Delta n$ will allow us to use the central limit,
while with a small $\Delta t$ we can resort to a continuous-time
approximation. At odds with the on-line MG, where characteristic times
are $\mathcal{O}(N)$ and this procedure is well defined as long as
$\Gamma\ll N$ \cite{mc}, in the present `batch' version this analysis is
in principle correct only for $\Gamma\ll 1$. However, it provides
useful insights whose validity for $\Gamma\gg 1$ can be checked
against numerical simulations.

Introducing the new variable $\widehat{y}_i(t)= \Gamma y_i(n)$ for
$t=n\Gamma$ and iterating (\ref{manna}) from time $n$ to time
$n+\Delta n$ we obtain
\begin{equation}\label{stra}
\widehat{y}_i(t+\Delta
t)-\widehat{y}_i(t)=\sum_{n=t/\Gamma}^{(t+\Delta t)/\Gamma}
\left\{-\ovl{\xi_i^\mu Q^\mu}
+\frac{1}{8}\left[(1-s_i(n))\zeta_{i1}(n)-(1+s_i(n))\zeta_{i2}(n)\right]
\right\}
\end{equation}
If $\Delta n$ is large enough, by the central limit theorem, we can
approximate the sum using the first two moments, whereas if $\Delta t$
is small enough, the change in $\widehat{y}_i(t)$ is small, which
means that $\widehat{y}_i(t)$ can be considered constant. The averages
over the fast spin variables can then be performed with the
``instantaneous'' distribution
\begin{equation}
{\rm Prob}\{s_i(n)=\pm 1\}\sim [1+e^{\mp\widehat{y}_i(t)}]^{-1}
\end{equation}
which depends on the ``slow'' variables $\widehat{y}_i(t)$. After
simple calculations we arrive, in the limit $\Delta t\to 0$, at the
continuous-time Langevin equation
\begin{gather}\label{langevino}
\partial_t\widehat{y}_i(t)=-\sum_{j=1}^N J_{ij}\tanh \widehat{y}_j -
\frac{\eta}{4}\tanh \widehat{y}_i+\epsilon_i(t)\\
\avg{\epsilon_i(t)\epsilon_j(t')}=\Gamma\sum_{k=1}^N
J_{ik}J_{jk}[1-(\tanh
\widehat{y}_k)^2]\delta(t-t')+\frac{\Gamma\Delta}{16}\delta_{i,j}
\delta(t-t')
\label{noise}
\end{gather}
where $J_{ij}=\ovl{\xi_i^\mu\xi_j^\mu}$. The case $\Gamma=0$ turns out
to differ from the case $\Gamma=0^+$ because the limits $t\to\infty$
and $\Gamma\to 0$ don't commute.

Let us first consider the case $\eta=0$. For $\Gamma=0^+$, when this
equation is exact, the dynamics becomes deterministic and the values
of $\widehat{y}_i$ (and hence of $\sigma^2$) can be obtained by
minimizing the function ${\cal H}_\eta$ given in (\ref{acca}). We were
unable to compute the stationary state of (\ref{langevino}) for
$\Gamma>0$ because, as in the case of the on-line MG for $H>0$, the
Fokker-Planck equation associated to it has no simple factorized
solutions (see \cite{mc}). Hence $\sigma^2$ depends on $\Gamma$ and on
$\Delta$ for all values of $c$. Such a dependence is weak for $c<c_c$
but it becomes very strong in the high density phase. In particular,
$\sigma^2$ increases with $\Gamma$, as in the MG.

\begin{figure}
\begin{center}
\includegraphics[width=7.5cm]{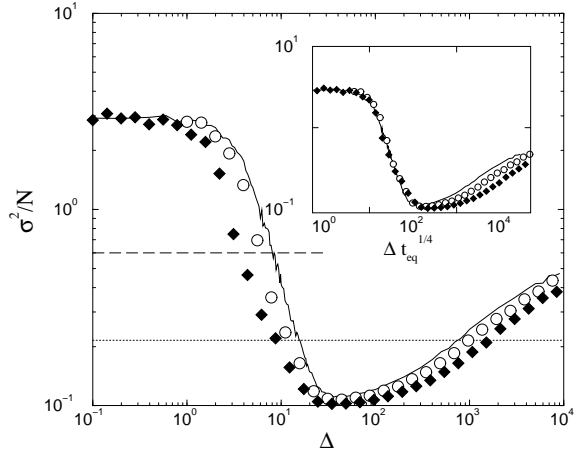}
\caption{\label{Delta} Behavior of $\sigma^2/N$ for a city of $P=200$
streets with $N=1024$ drivers as a function of the parameter $\Delta$
with $\eta=0$. The horizontal lines correspond to drivers with
$\Gamma=0$ (dashed) and $\Gamma=0^+$ (dotted). Results for adaptive
drivers with $\Gamma=\infty$ are shown for equilibration times $t_{\rm
eq}=100$ (full line), $400$ (open circles) and $1600$ (full
diamonds). In the inset, data are plotted versus $\Delta t_{\rm
eq}^{1/4}$.}
\end{center}
\end{figure}

When $\Delta=0$, the stationary state depends on the initial
conditions $\widehat{y}_i(0)$ for $c>c_c$: the larger the initial
spread, the smaller the value of $\sigma^2$ (see \cite{mc} for the
analogous phenomenon in the MG). For $\Delta>0$, the dependence on
initial conditions disappears and is replaced by a non-trivial
dynamical behavior (see Fig.~\ref{Delta}).
In the high density phase, where drivers would behave worse than
random with $\Delta=0$, global efficiency can improve beyond the
random threshold if $\Delta>0$. Taking averages after a fixed
equilibration time $t_{\rm eq}$, we find that $\sigma^2$ reaches, for
$\Delta\approx 40$, a minimum that is well below the value of
$\sigma^2$ for $\Gamma=0^+$ with the same homogeneous
initial conditions $y_i(0)=0$. For $\Delta\to\infty$ we recover the
behavior of random drivers. However, when we increase $t_{\rm eq}$,
the curve shifts to the left, showing that the system is not in a
steady state. Rescaling $\Delta$ by $t_{\rm eq}^{-1/4}$, the
decreasing part of the plot collapses, while the the rest of the curve
flattens. This suggests that the equilibrium value of $\sigma^2$ drops
suddenly as soon as $\Delta>0$. In order to understand this behavior,
it is useful to remind that large values of $\sigma^2$ are due to the
off-diagonal nature of the noise covariance (\ref{noise}), which is
responsible for a dynamic feedback effect, as shown in
\cite{cm03}. The presence of the diagonal noise term $\Delta$ destroys
the correlated fluctuations created by the first term of
(\ref{noise}), and removes the dependence on initial conditions
thereby allowing a slow diffusive dynamics toward the states of
maximal spread in the $y_i$ variables (i.e. with minimal $\sigma^2$)
which are consistent with the stationary state condition $H=0$. In
loose words: noise-corrupted information can avoid crowd effects when
the vehicle density is very high\footnote{We stress, however, that the
noise $\zeta_{ig}$ has to be uncorrelated across agents.}.

\begin{figure}
\begin{center}
\includegraphics[angle=-90,width=7cm]{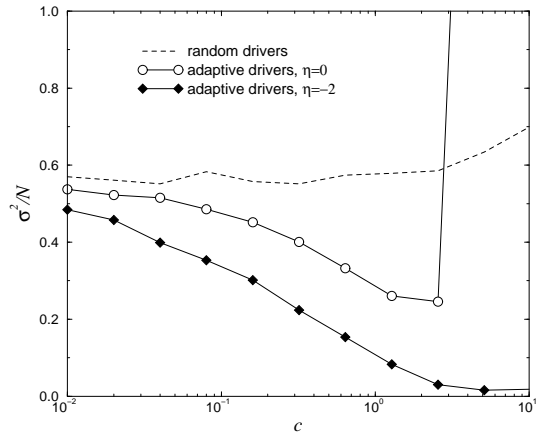}
\caption{\label{nash} Behavior of $\sigma^2/N$ for a city of $P=200$
streets with $\eta=-2$. Adaptive (resp. random) drivers have
$\Gamma=\infty$ (resp. $\Gamma=0$).}
\end{center}
\end{figure}

We finally come to the case $\eta\not =0$. If $\eta>0$ drivers
overestimate the profitability of routes they do not take. This
changes the phase transition into a smooth crossover, but the
qualitative picture remains the same. For $\eta<0$ instead the
situation changes radically. This effect has been studied in detail in
the MG \cite{cmz,dmm,dmh}. Based on these results, we argue that while
the stationary state is unique for $c<c_c(\eta)$, many stationary
states exist in the high density phase, each of which may be selected
by the dynamics, depending on initial conditions. This feature, which
corresponds to the occurrence of replica-symmetry breaking in the
minimization of (\ref{acca}), means dynamically that the system
acquires long term memory, as found in \cite{dmh} for the `na\"\i ve'
model with uncorrelated disorder and $\Delta=0$.  As $\eta\to 0^-$ we
find $c_c(\eta)\to c_c$, while for $\eta\to-2$ we find $c_c(\eta)\to
0$. $\eta<0$ implies that drivers overestimate the traffic on routes
they do not take. This means that they know that if they had actually
taken one of those route, the traffic there would have been slowed
down. At the value $\eta=-2$, drivers correctly account for this fact
and indeed the dynamics converges to an optimal solution, i.e. to a
Nash equilibrium in game theoretic terminology\footnote{A Nash
equilibrium here means that each driver takes the best route available
to him, given the choices of all other drivers.}. The stationary state
is characterized by no traffic fluctuations ($\sigma^2=H$) because
each driver selects one route and sticks to it. The value of
$\sigma^2$ for $\eta=-2$ is shown in Fig.~\ref{nash} as s function of
$c$.

In summary, we have introduced a highly simplified model for the
behavior of drivers in a city, with the aim of analyzing the emergent
collective behavior. The model is formally related to the `batch' MG
\cite{hc}, but displays features that are substantially different from
all previously studied variations. In absence of information noise,
inductive drivers turn out to behave better than random drivers for
low car densities, while at high densities the opposite
occurs. However, when $\Delta>0$ and/or $\eta<0$ inductive drivers can
be guided to a more efficient state, though the dynamics becomes
substantially more complex. Clearly, within its skinny definition, we
are unable to obtain a level of description as detailed as previous
approaches to urban traffic \cite{pr,cs}. However this model
highlights the relevance of the inductive behavior of drivers as
opposed to zero-intelligence random drivers, and hence the impact that
new technologies of information broadcasting can have on urban
traffic. Furthermore, the model is amenable to be made more realistic
(e.g. by considering routes with different $\ell$). This approach to
urban traffic can draw from the rich literature on the MG and its
variations, whose collective behavior has been clarified in detail, as
well as contribute to it.

\acknowledgments Work supported in part by the European Community's
Human Potential Programme under contract HPRN-CT-2002-00319, STIPCO,
and by the EU EXYSTENCE network. We thank G. Weisbuch and I. P\'erez
Castillo for useful discussions. The Abdus Salam ICTP is gratefully
acknowledged for hospitality and support.

\end{document}